\documentclass{elsart}

\usepackage{graphicx,psfrag,amsmath,amstext,latexsym}

\newcommand{\Order}[1]{\mathcal{O}#1}
\newcommand{\nn}{\nonumber}
\newcommand{\Mpi}{M_\pi}
\newcommand{\MK}{M_{\scriptscriptstyle{K}}}
\newcommand{\Mpic}{M_{\pi^{\scriptscriptstyle{+}}}}
\newcommand{\MKc}{M_{\scriptscriptstyle{K^+}}}
\newcommand{\ChPT}{CHPT\,}

\begin{document}
\begin{frontmatter}
 \hfill {\small UWThPh-2005-12}
\vspace*{1.5cm} 

\title{Isospin odd \boldmath{$\pi K$} scattering length}
  \author{J. Schweizer}
  \address{Institut f\"ur Theoretische Physik,\\
University of Vienna, A--1090 Vienna, Austria\\E-mail: julia.schweizer@univie.ac.at}

\begin{abstract}
We make use of the chiral two--loop representation of the $\pi K$ scattering amplitude  [J.~Bijnens, P.~Dhonte and P.~Talavera,
JHEP { 0405} (2004) 036] to investigate the isospin odd scattering length at next-to-next-to-leading order in the SU(3) expansion.
This scattering length is protected against contributions of $m_s$ in the chiral expansion, in the sense that the corrections to the current algebra result are of order $\Mpi^2$.
In view of the planned lifetime measurement on $\pi K$ atoms at CERN it is important to understand the size of these corrections.
\end{abstract}

\end{frontmatter}

\section{Introduction}
In the sixties and seventies a set of experiments was performed on $\pi K$ scattering \cite{Lang:1978fk}. 
To obtain predictions for the low--energy parameters, the measured $\pi K$ phases had to be extrapolated using dispersion relations and crossing symmetry 
\cite{Lang:1976ze}, since the region of interest is not directly accessible by scattering experiments. The most precise values for the $\pi K$ scattering lengths were obtained only recently from an analysis of Roy-Steiner equations \cite{Ananthanarayan:2000cp,Buettiker:2003pp}. 
Alternatively,  
particular combinations of $\pi K$ scattering lengths may be extracted from experiments on $\pi K$ atoms \cite{Deser:1954vq,Bilenky:1969zd,Schweizer}. The $\pi K$ atom decays due to the strong interactions into $\pi^0 K^0$ and a lifetime measurement will allow one to determine the isospin odd S-wave $\pi K$ scattering length $a_0^-=1/3(a^{1/2}_0-a^{3/2}_0)$. Such a measurement is planned at CERN \cite{futureDIRAC}.
Particularly interesting about the isospin odd $\pi K$ scattering length is that there exists a low--energy theorem due to Roessl \cite{Roessl:1999iu}. Based on SU(2) chiral perturbation theory (\ChPT) \cite{Roessl:1999iu,Weinberg:1978kz,Gasser:1983yg,Frink:2002ht},  where the strange quark mass is treated as a heavy partner, it is valid to all orders in powers of $m_s$. It states that Weinberg's current algebra result \cite{Weinberg:1966kf,griffith} receives corrections of order $\Mpi^2$ only,
 \begin{equation}
   a^-_0 = \frac{M_\pi \MK }{8\pi
   F_\pi^2(M_\pi+\MK)}\left\{1+\Order(\Mpi^2)\right\}.
   \label{eq: Roessl}
 \end{equation}
 Here $\Mpi$, $\MK$ and $F_\pi$ denote the physical meson masses and the physical pion decay constant. In view of this low--energy theorem, one would expect higher order corrections to the scattering length to be relatively small.  
 These days, the $\pi K$ scattering amplitude is available at next-to-next-to-leading order \cite{Bernard:1990kw,Kubis:2001bx,Nehme:2001wa,Bijnens:2004bu} in SU(3) \ChPT \cite{Gasser:1984gg}. The one--loop corrections \cite{Bernard:1990kw,Kubis:2001bx,Nehme:2001wa} to $a_0^-$ turn out as expected, they change the current algebra value at the $11\%$ percent level. Surprisingly, for the two--loop corrections  this seems not to be the case. According to the numerical study performed in Ref. \cite{Bijnens:2004bu}, the  scattering length $a_0^-$ receives at order $p^6$ a $14\%$ correction. The aim of the present article is to understand the nature of these rather substantial contributions at two--loop order. 
Other recent work on $\pi K$ scattering makes use of resonance chiral Lagrangian predictions \cite{Bernard:1991zc} together with resummations \cite{Jamin:2000wn}. There were also earlier attempts 
at unitarisation of current algebra for this process, see Ref. \cite{SaBorges:1994yy} and references therein. 

We use the chiral two--loop representation for the $\pi K$ amplitude \cite{Bijnens:2004bu} to investigate the order $p^6$ corrections to $a^-_0$. In Section \ref{sec: low cost}, we extract the contributions from the low--energy constants and determine the double chiral logs as well as the log$ \times L_i^r$ terms by means of the renormalization group equations for the renormalized coupling constants \cite{Bijnens:1999hw}. Further, we specify the 1-loop $\times L_i^r$ terms in an expansion in powers of $\Mpi/\MK$. The numerical analysis is carried out in Section \ref{sec: numerics} and the results for the partial two--loop contributions are collected in Table \ref{table: num delta4}.

\section{`Low cost' terms at two--loop order}
\label{sec: low cost}
The SU(3) chiral expansion of the isospin odd $\pi K$ scattering length looks as follows
\begin{equation}
a^-_0 = \frac{M_\pi\MK }{8\pi
   F_\pi^2(M_\pi+\MK)}\left\{1+\delta^{(2)}+\delta^{(4)}+\Order(p^6)\right\},
    \label{eq: am}
\end{equation}
  where $\Order(p^6) = \{ \hat{m}^3, \hat{m}^2 m_s, \hat{m} m_s^2\}$.
     The scattering length is expressed in terms of the physical meson masses $\Mpi$ and $\MK$ and the physical  pion decay constant $F_\pi$ \cite{Amoros:1999dp}.  The next-to-leading order contribution $\delta^{(2)}$ \cite{Kubis:2001bx,Nehme:2001wa} depends on one single low--energy constant $L_5^r$ \cite{Gasser:1984gg} only,
     \begin{align}
\delta^{(2)} &= \frac{\Mpi^2}{32\pi^2F_\pi^2}\left[ 256\pi^2 L_5^r-3\ln \frac{\MK^2}{\mu^2}-\frac{3(2\MK^2-\Mpi^2)}{\MK^2-\Mpi^2}\ln \frac{\Mpi^2}{\MK^2}\right.\nn\\
 &\quad\left.-\frac{4\MK^2-\Mpi^2}{2(\MK^2-\Mpi^2)}\ln \frac{4\MK^2-\Mpi^2}{3\Mpi^2}\right]+\frac{\Mpi \MK}{3F_\pi^2}\nn\\
 &\quad\times \left[ \bar{J}(s_{\rm thr},\MK^2,\tfrac{1}{3}(4\MK^2-\Mpi^2))-\bar{J}(u_{\rm thr},\MK^2,\tfrac{1}{3}(4\MK^2-\Mpi^2))\right]\!,
\label{eq: delta2}
\end{align}
where $s_{\rm thr} = (\Mpi+ \MK)^2$, $u_{\rm thr} =  (\MK-\Mpi)^2$ and the function $\bar{J}$ is defined as follows
\begin{align}
\bar{J}(p^2,m_1^2,m_2^2)&=J(p^2,m_1^2,m_2^2) -J(0,m_1^2,m_2^2),\nn\\
J(p^2,m_1^2,m_2^2) &= -i\int \frac{d^d q}{(2\pi)^d}(m_1^2-q^2)^{-1}(m_2^2-(p+q)^2)^{-1}.
\end{align}
Note that at the order considered  it makes a difference whether we represent  $\delta^{(2)}$ as a function of the physical pion, kaon and $\eta$ masses or express one of them through the other two\footnote{This will generate a correction proportional to $\Delta_{\rm GMO}\equiv (4\MK^2-\Mpi^2-3M_\eta^2)/(M_\eta^2-\Mpi^2)$ \cite{Gasser:1984gg} which contributes to $\delta^{(4)}$. }. In Eq. (\ref{eq: delta2}), we choose to describe $\delta^{(2)}$ in terms of the physical pion and kaon mass only,  because this ensures that both $\delta^{(2)}$ and $\delta^{(4)}$ are independently scale invariant. 

The two--loop order correction can be decomposed as
 \begin{equation}
\delta^{(4)} =  \delta^{(4)}_{L_i=C_i=0}+  \delta^{(4)}_{{\rm 1-loop}L_i} +  \delta^{(4)}_{L_i L_j} +\delta^{(4)}_{C_i} .
  \label{eq: delta4}
\end{equation}
 The first term contains the two--loop functions, the second one--loop functions with insertions of $\Order(p^4)$ coupling constants and the last two terms consist of counter term contributions.
Some of the two--loop functions in $\delta^{(4)}_{L_i=C_i=0}$ are very demanding to analyze analytically. For the moment, we thus restrict ourselves to the chiral double logs,
\begin{equation}
\delta^{(4)}_{L_i=C_i=0}=\delta^{(4)}_{{\rm log}^2}+\delta^{(4)}_{\rm rem},
 \label{eq: rem}
\end{equation}
 and neglect the remainder $\delta^{(4)}_{\rm rem}$ which is given numerically in Table \ref{table: num delta4}.
 In a first step, we extract the contributions from the $p^6$ low--energy constants ($C_i^r$) \cite{Bijnens:1999hw,Bijnens:1999sh} from the representation of the $\pi K$ scattering amplitude in Ref. \cite{Bijnens:2004bu},
\begin{align}
\delta^{(4)}_{C_i}&= \frac{16\Mpi^2}{F_\pi^2}\left[-2\MK^2 \left(C^r_1-2C^r_3-4C^r_4-C^r_{14}-C^r_{15}+2C^r_{22}\right.\right.\nonumber\\\
   &\quad \left. \left. -2C^r_{25}-C^r_{26}+2C^r_{29}\right)+\Mpi^2\left( C^r_{15}+2C^r_{17}\right)\right],
   \label{eq: Ci}
   \end{align}
 as well as products of two $p^4$ constants ($L_i^r\times L_j^r$),
  \begin{equation}
 \delta^{(4)}_{L_i L_j} =  \frac{64\Mpi^2 L_5^r}{F_\pi^4}\left\{\MK^2 \left[2(L_4^r-2L_6^r)-L_5^r\right]+\Mpi^2 \left[L_4^r-2L_6^r+2(L_5^r-L_8^r)\right]\right
  \}.
 \label{eq: Li2}
\end{equation}
 In order to determine the chiral double logs and the ${\rm log}\times L_i^r$ terms, we consider the renormalization group equations of  the renormalized order $p^4$ and $p^6$ low--energy constants \cite{Bijnens:1999hw},
\begin{equation}
\mu \frac{d L_i^r(\mu)}{d\mu}  = - \frac{1}{(4\pi)^2}\Gamma_i,\quad
\mu \frac{d C_i^r(\mu)}{d\mu}  = \frac{1}{(4\pi)^2}\left[ 2\Gamma_i^{(1)}+\Gamma_i^{(L)}(\mu)\right] .
\end{equation}
The coefficients $\Gamma_i^{(L)}$ are linear combinations of $p^4$ constants which satisfy the following differential equations,
 \begin{equation}
\mu \frac{d \Gamma_i^{(L)}(\mu)}{d\mu}  =- \frac{\Gamma_i^{(2)}}{8\pi^2},
\end{equation}
in accordance with Weinberg's consistency conditions \cite{Weinberg:1978kz}. The coefficients $\Gamma_i^{(1)}$, $\Gamma_i^{(2)}$ and $\Gamma_i^{(L)}(\mu)$ are listed in Table II of Ref. \cite{Bijnens:1999hw}. 
The solutions of the renormalization group equations read \cite{Jamin:2004re}
 \begin{align}
L_i^r(\mu) &= L_i^r(\mu_0)-\frac{\Gamma_i}{2}L(\mu/\mu_0),\nn \\
C_i^r(\mu) &= C_i^r(\mu_0)-\frac{1}{4}\Gamma_i^{(2)}L(\mu/\mu_0)^2+\frac{1}{2}\left[2\Gamma_i^{(1)}+\Gamma_i^{(L)}(\mu_0)\right] L(\mu/\mu_0),
 \label{eq: sRGE}
\end{align}
 with the chiral logarithm
 \begin{equation}
L(\mu/\mu_0) = \frac{1}{(4\pi)^2} \ln \frac{\mu^2}{\mu_0^2}.
\end{equation}
As a two--loop order quantity $\delta^{(4)}$  consists of
 \begin{equation}
\delta^{(4)} = \hat{a}(\mu)+\sum_i b_i C_i^r(\mu)+ \sum_{i,j}b_{ij} L_i^r(\mu)L_j^r(\mu),
\end{equation}
 where $\hat{a}(\mu)$ is scale dependent and contains one--loop functions with insertions of $p^4$ constants as well as two--loop functions.  In order to extract the double log and ${\rm log}\times L_i^r$ contributions from $\hat{a}(\mu)$, we insert the solutions for the renormalized coupling constants into the latter equation,
 \begin{align}
 \delta^{(4)} &= \hat{a}(\mu_0)+\sum_i b_i C_i^r(\mu_0)+ \sum_{i,j}b_{ij} L_i^r(\mu_0)L_j^r(\mu_0),\nn\\
 \hat{a}(\mu_0) &= \hat{a}(\mu) -\frac{1}{4}L(\mu/\mu_0)^2\left[b_i\Gamma_i^{(2)}-b_{ij}\Gamma_i \Gamma_j  \right]\nn\\
 &\quad+\frac{1}{2}L(\mu/\mu_0)\left[b_i\left(2\Gamma_i^{(1)}+\Gamma_i^{(L)}(\mu_0)\right)-2b_{ij}\Gamma_i L_j^r(\mu_0)\right].
 \end{align}
Now, the scale dependence of $\hat{a}(\mu_0)$ becomes apparent and we may read off the wanted log$^2$ and log$\times L_i^r$ terms. The solutions of the renormalization group equations thus allow us to determine the double log and ${\rm log}\times L_i^r$ contributions from Eqs. (\ref{eq: Ci}) and (\ref{eq: Li2}).  

 The double chiral logs (${\rm log}^2$) amount to
\begin{equation}
\delta^{(4)}_{{\rm log}^2} =\frac{\Mpi^2}{F_\pi^4}\left[\frac{37\MK^2}{8}+ \frac{59\Mpi^2}{24}\right]L(M_\chi/\mu)^2,
\label{eq: log2}
\end{equation}
while the single logarithms times $p^4$ constants (${\rm log}\times L_i^r$) yield
   \begin{align}
 \delta^{(4)}_{{\rm log} L_i} &=  \frac{-2\Mpi^2}{3F_\pi^4}\left\{ \MK^2\left[ 84L_1^r+114L_2^r+53L_3-96L_4^r-28L_5^r\right.\right.\nn\\
   &\quad \left.+48\left(3L_6^r+L_7+2L_8^r\right)\right]-\Mpi^2\left[12L_1^r+30L_2^r\right.\nonumber\\
   &\quad \left. \left. +19L_3-64L_5^r +24(2L_7+L_8^r)\right]\right\}L(M_\chi/\mu).
   \label{eq: logLi}
\end{align}
 Here $M_\chi$ stands for a characteristic meson mass. 

In the remaining part of this section, we investigate Roessl's low--energy theorem \cite{Roessl:1999iu} at next-to-next-to-leading order in SU(3) \ChPT. More precisely, we specify the order $\Mpi^2$ and order $\Mpi^4$ corrections to Eq. (\ref{eq: Roessl}). To approach the SU(2) chiral expansion, we regard the kaon mass as heavy and expand $a^-_0$ in powers of $ M_\pi/M_K$,
 \begin{equation}
   a^-_0 = \frac{M_\pi \MK }{8\pi
   F_\pi^2(M_\pi+\MK)}\left\{1+M_\pi^2c_{2}+M_\pi^4c_{4}+\Order(M_\pi^6)\right\}.
 \label{eq: amSU2}
 \end{equation}
 Again, the quantities $\Mpi$, $\MK$ and $F_\pi$ stand for the physical masses and the physical pion decay constant \cite{Amoros:1999dp}. 
  At next-to-leading order in SU(3) \ChPT, the coefficient $c_{2}$ depends on $L_5^r$ \cite{Kubis:2001bx,Nehme:2001wa},
\begin{align}
c_{2} \mid_{\rm 1-loop}&=\frac{1}{F_\pi^2}\left\{8L^r_5-\frac{1}{32\pi^2}\left[3 \ln \frac{\MK^2}{\mu^2}+4\ln \frac{\Mpi^2}{\MK^2}\right]\right.\nn\\
&\quad\left.+\frac{1}{144\pi^2}\left[ -12+10\sqrt{2}\arctan{\sqrt{2}}-7\ln \frac{4}{3}\right]\right\},
\label{eq: c2p4}
 \end{align}
while the one--loop contributions to $c_{4}$ do not contain any low--energy constants and can safely be neglected numerically. At  next-to-next-to-leading order in the chiral SU(3) expansion, the contributions from counter terms, double chiral logs and log$\times L_i^r$ terms to the coefficients $c_2$ and $c_4$ are specified  in Eqs. (\ref{eq: Ci}), (\ref{eq: Li2}), (\ref{eq: log2}) and (\ref{eq: logLi}). In addition, we list the expansion of the one--loop functions with insertions of $p^4$ couplings in powers of $\Mpi/\MK$. We have
\begin{align}
c_2\mid _{{\rm 1-loop}L_i}&=\frac{\MK^2}{12\pi^2F_\pi^4}\left\{-\frac{1}{2}\large[84 L_1^r+114L_2^r+53L_3-96L_4^r-28L_5^r\right.\nn\\
&\quad\left.+48\left(3L_6^r+L_7+2L_8^r\right)\right]\ln \frac{\MK^2}{\mu^2}\nn\\
&\quad-\frac{4}{27}L_3\left[ 56\sqrt{2}\arctan\sqrt{2}-5\ln\frac{4}{3}\right]\nn\\
&\quad-\frac{1}{3}\left[L_5^r-6(2L_7+L_8^r)\right]\left[ 13\sqrt{2}\arctan\sqrt{2}+2\ln\frac{4}{3}\right]+93L_1^r\nn\\
&\quad+\frac{189}{2}L_2^r+\frac{2045}{36}L_3-16\left[L_5^r+6(L_4^r-L_6^r+L_7)\right]\bigg\},
\label{eq: c2}
\end{align}
and
\begin{align}
c_4\mid _{{\rm 1-loop}L_i}&=\frac{1}{8\pi^2F_\pi^4}\left\{\frac{1}{3}\left[12L_1^r+30L_2^r+19L_3-64L_5^r \right.\right.
\nn\\
&\quad\left.+24(2L_7+L_8^r)\right]\ln \frac{\MK^2}{\mu^2} +4\large[8L_1^r+12L_2^r+6L_3-8L_4^r\nn\\
&\quad-9L_5^r+6(2L_6^r+L_8^r)\large]\ln \frac{\Mpi^2}{\MK^2}-\frac{\sqrt{2}}{8}\bigg[\frac{1840}{81}L_3-\frac{1415}{18}L_5^r\nn\\
&\quad+45(2L_7+L_8^r)\bigg]\arctan \sqrt{2}+\frac{4}{9}\bigg[\frac{2}{9}L_3-17L_5^r\nn\\
&\quad+18(2L_7+L_8^r)\bigg]\ln\frac{4}{3}-\frac{1}{4}\left[8L_1^r+4L_2^r-\frac{410}{27}L_3\right.\nn\\
&\quad\left.+\frac{323}{6}L_5^r-67(2L_7+L_8^r)\right]\bigg\},
\label{eq: c4}
\end{align}
where we have checked that the log$\times L_i^r$ terms agree with Eq. (\ref{eq: logLi}). Here both the contributions to $\Mpi^2c_2$ and $\Mpi^4c_4$ are numerically sizeable, see Table \ref{table: num delta4}. 

 \section{Numerical analysis} 
 \label{sec: numerics}
 
  \begin{table}[t]
    \begin{center}
 \begin{tabular}{lllllll}\hline\rule{0mm}{4mm}
& CA & SU(2) \cite{Roessl:1999iu}& $p^4$ SU(3) \cite{Kubis:2001bx} & $p^6$ SU(3) \cite{Bijnens:2004bu} & Ref. \cite{Buettiker:2003pp}\vspace*{0.1em} \\ \hline\hline\rule{0mm}{5mm}
\hspace*{-0.2em}$\Mpi a_0^-$ & $0.071$& $0.077\pm0.003^*$& $0.0793\pm0.0006$ &  $0.089$ &$0.090\pm0.005$\vspace*{0.1em}\\ 
\hline
\end{tabular}\vspace*{0.4em}
   \caption{Isospin odd scattering length $a_0^-$: CA current algebra value, SU(2)  
prediction \cite{Roessl:1999iu}, chiral SU(3) prediction at order $p^4$ \cite{Kubis:2001bx} and order $p^6$ \cite{Bijnens:2004bu}, dispersive analysis from Roy-Steiner equations \cite{Buettiker:2003pp}.
*Note that in Ref. \cite{Roessl:1999iu}  $\Mpi = 137.5$ MeV and $\MK=495.5$ MeV, while all other references use $\Mpi \doteq \Mpic$ and $\MK\doteq\MKc$ for the pion and kaon masses in the isospin symmetry limit.}
\label{table: num am}
\end{center}
\end{table}
 In the following, we present the numerical results for the partial $p^6$ corrections to $\delta^{(4)}$.
  The pion and kaon mass in the isospin symmetry limit are identified with their charged masses $\Mpi \doteq \Mpic$ and $\MK\doteq\MKc$. To be consistent with the numerical analysis performed in Ref. \cite{Bijnens:2004bu}, we use for the pion decay constant\footnote{Recently, a new value was obtained $F_\pi = 92.2\pm0.2$ MeV \cite{Descotes-Genon:2005pw}.} $F_\pi = 92.4$ MeV. In Table \ref{table: num am}, we list the various numerical results for $a^-_0$ available in the literature. The first row contains the current algebra value, the next number is the SU(2) prediction at next-to-leading order \cite{Roessl:1999iu}, row three and four display the order $p^4$ \cite{Kubis:2001bx} and order $p^6$ \cite{Bijnens:2004bu} SU(3) predictions and the last value is based on a phenomenological analysis from Roy-Steiner equations \cite{Buettiker:2003pp}. As can be read off, the SU(3) prediction at order $p^6$ is in good agreement with the Roy-Steiner value. The SU(3) chiral expansion of the scattering length $a_0^-$ looks as follows
  \begin{align}
 \frac{8\pi
   F_\pi^2(M_\pi+\MK)}{\MK M_\pi}a^-_0 &=1+\delta^{(2)}+\delta^{(4)}+\cdots \nonumber \\
   &= 1+0.11+0.14+\cdots
     \label{eq: SU3exp}
\end{align}
The one--loop contribution $\delta^{(2)}$ changes the current algebra result at the $11\%$ level, while the two--loop contributions $\delta^{(4)}$ amount to a $14\%$ correction. The aim was to understand this rather large order $p^6$ correction and our insights are collected in Table \ref{table: num delta4} which contains a splitting up of the various contributions at two--loop order. 

\begin{table}[t]
 \begin{center}
 \begin{tabular}{llllll}\hline\rule{0mm}{4mm}
$a$  &$\delta_a^{(4)}$&  $\Mpi^2 c_{2}\mid_a $ & $\Mpi^4 c_{4} \mid_a$ & $\beta\mid_a$ \vspace*{0.1em}\\ \hline\hline\rule{0mm}{6mm}
$L_i = C_i = 0$ & $0.05^\star$& -&-  &- \\ \rule{0mm}{5mm}
${\rm log}^2$ &  $0.010$ & $0.010$& $0.0004$ & $3.7$ \\\rule{0mm}{5mm}
1-loop$ \,L_i$ &  $0.013$ &  $0.007$&  $0.006$& $2.9$\\\rule{0mm}{5mm}
$L_i  L_j$ & $-0.004$ & $-0.004$&$0.0002$ & $-1.5$\\\rule{0mm}{5mm}
$C_i$& $0.08^\dagger$& $0.08$& $0$&$30.6$\\\rule{0mm}{5mm}
rem &$0.04$ &- & -&-
\vspace{0.5em} \\ 
\hline
\end{tabular}
\vspace*{0.4em}
\caption{Numerical results for the $p^6$ contributions at the scale $\mu = 770$ MeV: 
$^\star$ pure loop contributions and $^\dagger$ resonance estimate are taken from Ref. \cite{Bijnens:2004bu}. The notation is understood as in Eq. (\ref{eq: delta4}). For instance the contributions of the 1-loop$\times L_i^r$  terms to $\delta^{(4)}$ is given by $\delta^{(4)}_{{\rm 1-loop}L_i} = 0.013$.
}

\label{table: num delta4}
\end{center}

\end{table}

For the low--energy constants $L_i^r$ at the scale $\mu = 770$ MeV ($M_\rho$),  we use fit 10 of Ref. \cite{Amoros:2001cp}.
 The double chiral logs are evaluated for a characteristic meson mass\footnote{The choice $M_\chi = \sqrt{ \Mpi \MK}$ leads to an unnatural large number for the double logs $\delta^{(4)}_{{\rm log}^2}=0.058$, to be compared with the full pure loop corrections $\delta^{(4)}_{L_i = C_i = 0} = 0.05$ \cite{Bijnens:2004bu}.
 For $M_\chi = \Mpi$ the value becomes even more unreasonable.} $M_\chi = \MK$  and the size of the remainder $\delta^{(4)}_{\rm rem}$ is estimated by the use of Eq. (\ref{eq: rem}). Row two and three of Table \ref{table: num delta4} contain the partial order $p^6$ corrections to the coefficients $c_2$ and $c_4$, respectively. Note that for the double chiral logs as well as for the products of $p^4$ constants their contribution to $c_4$ can be neglected while for the one--loop functions with insertions of $L_i^r$'s, both $\Mpi^2 c_2$ and $\Mpi^4 c_4$ are numerically sizeable. The enhancement of the coefficient $c_4$ is mainly due the contributions proportional to $\ln \Mpi/\MK$, see Eq.  (\ref{eq: c4}). 
 
As one can read off from Table \ref{table: num delta4}, more than half of the contributions to $\delta^{(4)}=0.14$ stem from the resonance estimate for the $p^6$ constants which includes effects of the lowest-lying vector and scalar resonances \cite{Bijnens:2004bu}. 
We checked that with this procedure the meson resonance exchange contributions to $C_{15}^r$ and $C_{17}^r$ vanish which implies that $c_4\mid_{C_i}$ is equal to zero. Further, for the combination of $p^6$ constants occurring in $c_2\mid_{C_i}$, the contributions from scalar resonances do not play a dominant role: They amount to $0.03$ of the $0.08$ generated by the $C_i^r$'s in total.
 It would be instructive to see whether these features persist in an improved estimate for the $p^6$ constants which respects the constraints that follow by imposing the proper asymptotic behaviour for massless QCD \cite{Ecker:1988te}.
 
  \begin{table}[t]
\begin{center}
 \begin{tabular}{lllllll}\hline
& ${L_i = C_i = 0}$ &{1-loop}\,$L_i$ & $ {L_iL_j}$&${C_i}$\vspace*{0.1em}\\ \hline\hline\rule{0mm}{6mm}
$\Delta \delta_a^{(4)}$ & $-0.03$& $0.02$ & $-0.01$& $0.02$\\
\hline
\end{tabular}
\vspace*{0.4em}
\caption{Variations of the partial $p^6$ contributions to $\delta^{(4)}$ for $M_\eta \le \mu \le 770\, {\rm MeV} \,(M_\rho)$. More precisely, we display the difference $\Delta \delta_a^{(4)} = \delta^{(4)}_a \mid_{\mu =M_\eta}-\delta^{(4)}_a \mid_{\mu = M_\rho}$. For the notation, see Table \ref{table: num delta4}.}

\label{table: num scale}
\end{center}
\end{table}

The splitting of the order $p^6$ contributions in Table \ref{table: num delta4} is scale dependent. Table \ref{table: num scale} displays the scale dependence of the various contributions to $\delta^{(4)}$.  The values for the 1-loop$\times L_i^r$, $L_i^r\times L_j^r$ and $C_i^r$ terms at the scales $\mu = 770$ MeV and $\mu = M_\eta$ allow us to read off the scale dependence of the pure loop contributions $\delta^{(4)}_{L_i=C_i =0}$.
 
Finally, we sum up the various SU(3) one- and two--loop contributions to $c_2$ and $c_4$ and get for the expansion of $a_0^-$ in powers of $\Mpi/\MK$,
 \begin{align}
\frac{8\pi F_\pi^2(\Mpi+\MK)}{\Mpi\MK}a^-_0 &=1+\Mpi^2c_2+\Mpi^4 c_4+\cdots \nonumber \\
   &= 1+0.2+0.01+\delta^{(4)}_{\rm rem} + \cdots
   \label{eq: SU2exp}
\end{align}
Note that this decomposition is valid up to the contribution of $\delta^{(4)}_{\rm rem}=0.04$ only. 
Compared to the chiral SU(3) expansion in Eq. (\ref{eq: SU3exp}), the series in $\Mpi/\MK$ converges much more rapidly. 
The correction $\Mpi^2 c_2$ consists of 
\begin{equation}
 \Mpi^2 c_2 = \frac{\Mpi^2}{(4\pi F_\pi)^2} \left[ \alpha+\frac{\MK^2}{(4\pi F_\pi)^2}\beta+\cdots \right],
 \label{eq: alpha}
\end{equation}
where the coefficients $\alpha$ and $\beta$ contain the one--loop and two--loop contributions, respectively. Numerically, we have $\alpha = 7.6 $, where the dominant part stems from the term proportional to $\ln \Mpi/\MK$ in Eq. (\ref{eq: c2p4}). The contributions from double logs, 1-loop$\times L_i$ terms and $p^6$ constants to $\beta$ are listed in Table \ref{table: num delta4}. Here the bulk part comes from the resonance estimate for the $p^6$ constants \cite{Bijnens:2004bu}.

\section{Conclusions}
In the present work, we used the chiral two--loop representation for the $\pi K$ amplitude available in the literature \cite{Bijnens:2004bu} to investigate the isospin odd S-wave scattering length $a_0^-$. This scattering length 
differs from other low--energy parameters in $\pi K$ scattering 
in the sense that contributions of $m_s$ in the chiral expansion are suppressed by powers of $\hat{m}$.
Based on SU(2) CHPT \cite{Roessl:1999iu}, there exists a low--energy theorem (\ref{eq: Roessl}) which states that the current algebra result for $a_0^-$ receives corrections of order $\Mpi^2$ only. 
It was therefore expected that the one--loop result \cite{Bernard:1990kw,Kubis:2001bx,Nehme:2001wa} in SU(3) CHPT represents a decent estimate for the scattering length. However, the dispersive analysis from Roy-Steiner equations \cite{Buettiker:2003pp} and the chiral two--loop calculation \cite{Bijnens:2004bu} are not in agreement with this expectation. In fact, the numerical analysis performed in Ref. \cite{Bijnens:2004bu} showed that the two--loop order corrections to $a_0^-$ are of the same order of magnitude as the one--loop contributions.

In order to understand this rather substantial next-to-next-to-leading order correction, we determined analytically the contributions containing $p^6$ constants (\ref{eq: Ci}), products of two $p^4$ constants (\ref{eq: Li2}), double chiral logs (\ref{eq: log2}) and single logarithms times $p^4$ constants (\ref{eq: logLi}). We further expanded the one--loop functions with insertions of $p^4$ constants in powers of $\Mpi/\MK$, see Eqs. (\ref{eq: c2}) and (\ref{eq: c4}). The expansion of the pure two--loop functions in powers of $\Mpi/\MK$ was beyond the scope of this work. The numerical values of the partial $p^6$ contributions are collected in Table \ref{table: num delta4}. 

In the remaining part of this work, we investigated the low--energy theorem for $a_0^-$ at next-to-next-to-leading order in the SU(3) expansion. While it is true that the corrections are of order $\Mpi^2$, the chiral expansion of the accompanying coefficient proceeds in powers of $\MK$ and is not protected against sizeable contributions.  
At two--loop accuracy in the SU(3) expansion, the order $\Mpi^2$ correction roughly amounts to about $20\%$, see Eq. (\ref{eq: SU2exp}). Note that this number depends on the resonance estimate  \cite{Bijnens:2004bu} for the $p^6$ constants. If we compare this result with Roessl's value \cite{Roessl:1999iu}, the SU(2) prediction for the scattering length $a^-_0$ seems to be underestimated. At first surprisingly, we have to keep in mind that the numerical estimates for the low--energy constants in SU(2) \ChPT  were obtained through matching the scattering amplitude with the corresponding SU(3) \ChPT result at one--loop order. 
It would be very interesting to estimate these low--energy constants using a resonance saturation approach in the context of SU(2) CHPT with strangeness number $1$.
\section*{Acknowledgments}
It is a pleasure to thank J. Gasser for many helpful comments and discussions. Further, I thank J. Bijnens, G. Ecker, R. Kaiser and P. Talavera for useful remarks and discussions, and J. Bijnens for providing me with fortran programs for the two--loop $\pi K$ scattering amplitude.
This work was supported by the Swiss National Science Foundation.

\end{document}